\documentclass[aps,pra,twocolumn,amsmath,amssymb,showpacs,superscriptaddress]{revtex4-1}
\usepackage{color}
\usepackage{textcomp} 
\usepackage{mathrsfs,amsmath}
\usepackage{graphicx}
\usepackage{dcolumn}
\usepackage[mathlines]{lineno}
\usepackage{hyperref}
\usepackage{bm}
\usepackage{epstopdf}
\usepackage{soul}
\usepackage{epsfig}
\usepackage{bbold}

\usepackage{booktabs}

\usepackage{changes}
\usepackage{lipsum}
\setremarkmarkup{(#2)}

\newcommand{\bra}[1]{\ensuremath{\left\langle#1\right|}}
\newcommand{\ket}[1]{\ensuremath{\left|#1\right\rangle}}

\usepackage{color}

\begin{document}

\title{Recovery of local entanglement in self-healing vector vortex Bessel beams}

\author{Eileen Otte}
\affiliation{Institute of Applied Physics, University of Muenster, Corrensstr. 2/4, D-48149 Muenster, Germany}
\author{Isaac Nape}
\affiliation{School of Physics, University of the Witwatersrand, Private Bag 3, Wits 2050, South Africa}
\author{Carmelo Rosales-Guzm\'an}
\email[Corresponding author: ]{carmelo.rosalesguzman@wits.ac.za}
\affiliation{School of Physics, University of the Witwatersrand, Private Bag 3, Wits 2050, South Africa}
\author{Adam Vall\'es}
\affiliation{School of Physics, University of the Witwatersrand, Private Bag 3, Wits 2050, South Africa}
\author{Cornelia Denz}
\affiliation{Institute of Applied Physics, University of Muenster, Corrensstr. 2/4, D-48149 Muenster, Germany}
\author{Andrew~Forbes}
\affiliation{School of Physics, University of the Witwatersrand, Private Bag 3, Wits 2050, South Africa}
\date{\today}

\begin{abstract}
\noindent 
One of the most captivating properties of diffraction-free optical fields is their ability to reconstruct upon propagation in the presence of an obstacle both, classically and in the quantum regime. Here we demonstrate that the local entanglement, or non-separability, between the spatial and polarisation degrees of freedom also experience self-healing. We measured and quantified the degree of non-separability between the two degrees of freedom when propagating behind various obstructions, which were generated digitally. Experimental results show that even though the degree of non-separability reduces after the obstruction, it recovers to its maximum value within the classical self-healing distance. To confirm our findings, we performed a Clauser-Horne-Shimony-Holt Bell-like inequality measurement, proving the self-reconstruction of non-separability. These results indicate that local entanglement between internal degrees of freedom of a photon, can be recovered by suitable choice of the enveloping wave function.
\end{abstract}

\maketitle
\section{Introduction}

Self-healing is one of the most fascinating properties of diffraction-free optical fields\,\cite{Durnin1987}. These fields have the ability to reconstruct if they are partially disturbed by an obstruction placed in their propagation path. Diffraction-free beams have found applications in fields such as imaging~\cite{Fahrbach2010,Planchon2011,Fahrbach2012}, optical trapping~\cite{McGloin2005,McGloin2003,ARLT2001,Volke2002}, laser material processing~\cite{Dubey2008}, amongst many others. Arguably, the most  well-known propagation invariant (self-healing) fields are Bessel modes of light, first introduced in 1987 by J. Durin\,\cite{Durnin1987,Durnin1987b}. However, the self-healing property is not limited to so called non-diffracting beams, but also appears in helico-conical\,\cite{Hermosa2013}, caustic, or self-similar fields, namely Airy\,\cite{Broky2008}, Pearcey\,\cite{Ring2012}, Laguerre-Gaussian\,\cite{Bouchal2002,Mendoza2015} and even standard Gaussian beams\,\cite{Aiello2017}. Furthermore, within the last years, it has been shown that self-healing can also be observed at the quantum level, for example, McLaren {\it et al.} demonstrated experimentally the self-reconstruction of quantum entanglement\,\cite{McLaren2014}. Importantly, self-healing is not only an attribute of scalar fields but it can also apply to beams with spatially variant polarization \cite{Milione2015d,Wu2014,Li2017}. \\
Bessel beams also appear as complex vector light fields, where polarisation and spatial shape can be coupled in a non-separable way\,\cite{Zhan2009,Rosales2017,Otte2018}. This property has fueled a wide variety of applications, from industrial processes, such as, drilling or cutting\,\cite{Meier2007,Niziev1999,Dubey2008}, to optical trapping\,\cite{Donato2012,Zhao2005,Zhan2004,Roxworthy2010,Kozawa2010,Huang2012}, high resolution microscopy\,\cite{Torok2004}, quantum and classical communication\,\cite{Zhan2002,Ndagano2017,Ndagano2018}, amongst many others. Controversially, such non-separable states of classical light are sometimes referred to as classically or non-quantum entangled \,\cite{simon2010nonquantum}. This stems from the fact that the quintessential property of quantum entanglement is non-separability, which is not limited to quantum systems. Indeed, the equivalence has been shown to be more than just a mathematical construct \cite{Ndagano2017}.  While such classical non-separable fields do not exhibit non-locality, they manifest all other properties of local entangled states.\\
Here, we demonstrate that the decay in such local entanglement after an obstruction can be counteracted if the carried field is Bessel. We create higher-order vector Bessel beams that are non-separable in orbital angular momentum (the azimuthal component of the spatial mode) and polarisation, and show that self-healing also comprises the non-separability of the beams. This is at a first glance surprising since self-healing is traditionally attributed to the radial component of the spatial mode, which in our field is entirely separable. In order to demonstrate the far-reaching concept of self-healing, we unambiguously quantify the degree of non-separability in different scenarios ranging from fully to partially reconstructed fields by performing a state tomography on the classical field \cite{McLaren2015,Ndagano2016}. We show both theoretically and experimentally that even though the non-separability reduces after the obstruction, it recovers again upon propagation, proportionally to the level of self-reconstruction. Further, we confirm our findings by a Bell-like inequality measurement\,\cite{bell1964js} in its most commonly used version for optics, namely, the Clauser-Horne-Shimony-Holt (CHSH) inequality \,\cite{clauser1969}, confirming that the non-separability of vector Bessel beams also features self-healing properties. Although our tests are exerted on purely classical fields, the results are expected to be identical for the local entanglement of internal degrees of freedom of a single photon, and may be beneficial where such entanglement preservation is needed, e.g., transporting single photons through nano-apertures for plasmonic interactions.

\begin{figure*}[ht]
\centering
\def\svgwidth{0.9\linewidth}\sffamily
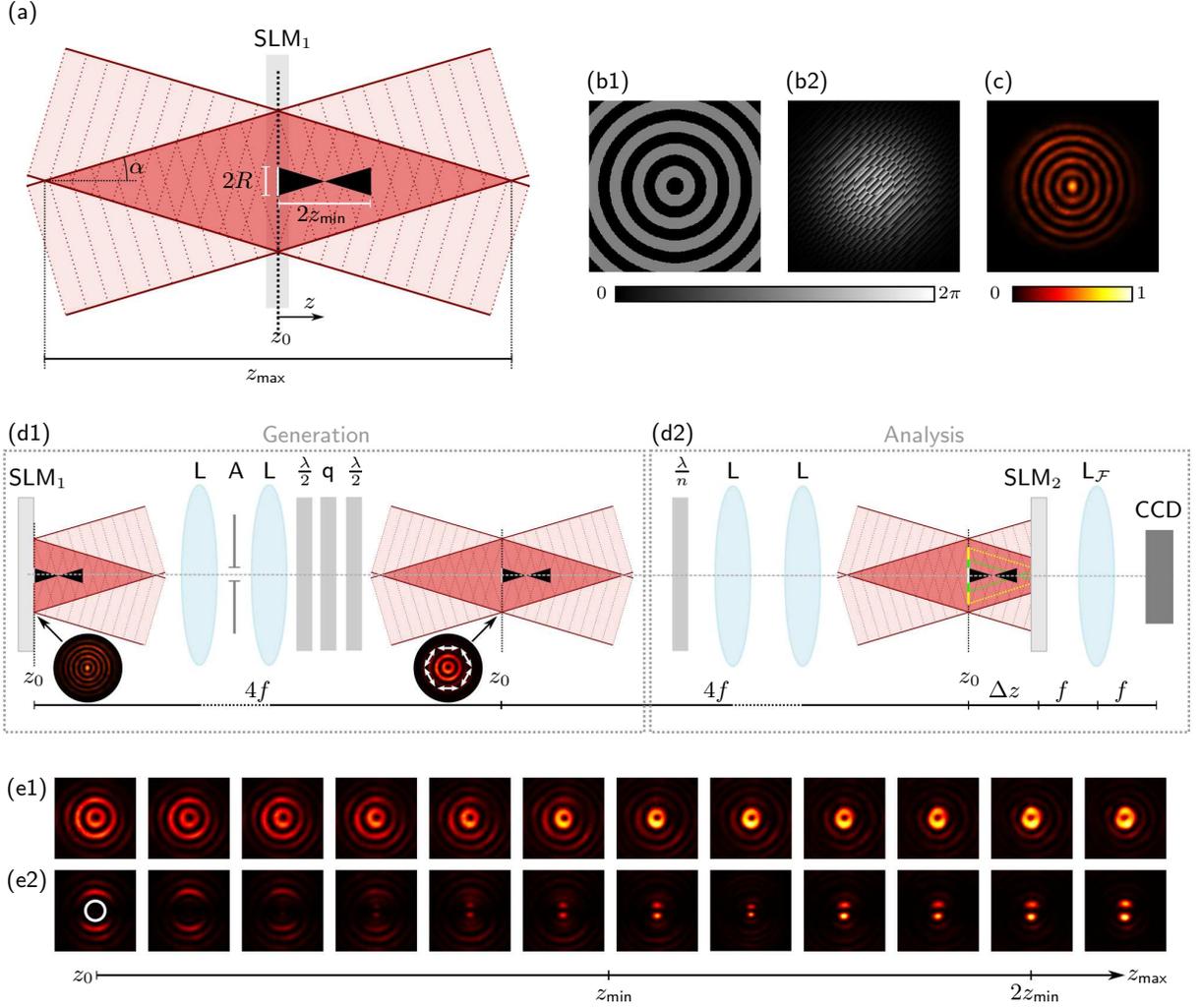  
\caption{(a) Formation of Bessel beams by off-axis interfering plane waves. If obstacles are included (radius $R$) a shadow region is formed. Scalar BG modes (intensity profile at $z_0$ in (c)) are realized by applying the binary Bessel function (b1) in combination with a blazed grating and Gaussian aperture (b2). (d) Concept of the realization and analysis of non-separable vBG modes (SLM: spatial light modulator; L$_{(\mathcal{F})}$: (Fourier) lens with focal distance $f$; A: aperture; $\lambda/n$, $n=\lbrace2,\,4\rbrace$: wave plates; q: $q$-plate; CCD: camera). (e) Propagation behavior of azimuthal vBG mode (e1) without obstruction and (e2) obstructed by on-axis absorbing object with $R=200\,\mbox{\textmu m}$, indicated by white circle.\label{fig:Concept}} 
\end{figure*}

\section{Vector Bessel modes} \label{sec:VectorBesselModes}
\subsection{Bessel-Gaussian beams}
\noindent Over finite distances, a valid approximation of Bessel beams is given by the so called Bessel-Gaussian (BG) modes\,\cite{Gori1987}. Besides other properties, these BG fields have the same ability of self-reconstruction in amplitude and phase\,\cite{McGloin2003,Litvin2009}. In polar coordinates $(r,\varphi, z)$, BG modes are defined as
\begin{align}
E_{\ell}^{\text{BG}}(r,\varphi, z) = &\sqrt{\frac{2}{\pi}}J_{\ell}\left(\frac{z_R k_r r}{z_R-\text{i}z}\right)\exp\left(\text{i}\ell\varphi-\text{i}k_z z\right) \nonumber \\
&\cdot \exp\left(\frac{\text{i}k_r^2z w_0-2kr^2}{4(z_R-\text{i}z)}\right),
\label{eq:BGmodes}
\end{align}
whereby $\ell$ represents the azimuthal index (topological charge), and $k_r$ and $k_z$ are the radial and longitudinal wave numbers, respectively. Further, $J_{\ell}(\cdot)$ defines the Bessel function, whereas the Gaussian information is encoded in the last factor with the initial beam waist $w_0$ of the Gaussian profile and the Rayleigh range $z_R = \pi w_0^2/\lambda$, $\lambda$ being the wavelength. The finite propagation distance of BG modes (``non-diffracting length") is limited by $z_{\text{max}}$. This distance describes the length of a rhombus-shaped region created by the superposition of plane waves with wave vectors lying on a cone described by the angle $\alpha=k_r/k$ (wave number $k=2\pi/\lambda$)\,\cite{Gori1987}, as indicated in Fig. \ref{fig:Concept}(a). The center of the rhombus-shaped region is positioned at $z_0$. For small $\alpha$, i.e., $\sin \alpha \approx \alpha$, the non-diffracting distance is given by $z_{\text{max}} =2\pi w_0/\lambda k_r$ \cite{McGloin2005}. If an obstruction is included within the non-diffracting distance, a shadow region is formed of length $z_{\text{min}} \approx R/\alpha \approx \frac{2\pi R}{k_r \lambda}$ \cite{Bouchal1998} (Fig.~\ref{fig:Concept}(a)). Here, $R$ describes the radius of the obstruction. After this distance $z_{\text{min}}$, the beam starts to recover due to the plane waves passing the obstruction \cite{McGloin2003,Litvin2009}. A fully reconstructed BG beam will be observed at $2z_{\text{min}}$, as visualised in Fig.~\ref{fig:Concept}(a).

\subsection{Realization of obstructed BG modes}
\noindent An established tool for the realization of complex beams are spatial light modulators (SLMs). These modulators allow for an on-demand dynamic modulation of structured beams by computer generated holograms \cite{SPIEbook}. For the formation of BG modes, we choose a binary Bessel function as phase-only hologram, defined by the transmission function
\begin{equation}
T(r,\varphi) = \text{sign}\lbrace J_{\ell}(k_r r)\rbrace \exp (\text{i}\ell\varphi), \label{eq:BinaryBessel}
\end{equation}
with the sign function $\text{sign}\lbrace\cdot \rbrace$ \cite{Turunen1988,Cottrell2007}. This approach has the advantage of generating a BG beam immediately after the SLM. An example of this function is shown in Fig. \ref{fig:Concept}(b1). Note that for encoding this hologram we use a blazed grating (see Fig.~\ref{fig:Concept}(b2)), so that the desired beam is generated in the first diffraction order of the grating\,\cite{Davis1999}.\\
Here, we set $k_r = 18\,\mbox{rad\,mm}^{-1}$ and $\ell =0$ for the fundamental Bessel mode. Furthermore, we multiply the hologram by a Gaussian aperture function for the realization of a Gaussian envelope with $w_0 = 0.89\,\mbox{mm}$ (see Eq. (\ref{eq:BGmodes}), Fig. \ref{fig:Concept}(b2)). These settings result in a BG beam with $z_{\text{max}} = 49.16\,\mbox{cm}$ for a wavelength of $\lambda = 633\,\mbox{nm}$, whose intensity profile in the $z_0$-plane is depicted in Fig.~\ref{fig:Concept}(c).\\
Beyond the generation of BG modes, the SLM can also be used for the realization of obstructions within the $z_0$-plane (see Fig. \ref{fig:Concept}(d1)): Absorbing obstacles are created by including a circular central cut in the hologram, such that within this area no blazed grating is applied. This means, the respective information of the BG mode is deleted in the first diffraction order. Furthermore, phase obstructions can be realized by adding the chosen phase object to the hologram. Hence, this artificial generation of obstructions facilitates the realisation of any chosen kind of obstacle of defined radius $R$ in the $z_0$-plane. Moreover, the relation $z_{\text{min}} \approx \frac{2\pi R}{k_r \lambda}$ shows that a decrease in the radius R of the circular obstruction at $z_0$ results in a decrease in the length $z_{\text{min}}$ of the shadow region, which is equivalent to moving the detection plane in $\pm z$-direction (see Fig. \ref{fig:Concept}(d2)), in order to analyse the evolution from a partially to a fully reconstructed beam.

\subsection{Self-healing vector Bessel modes}
\begin{figure}[h]
\centering
\def\svgwidth{1.0\linewidth}\sffamily
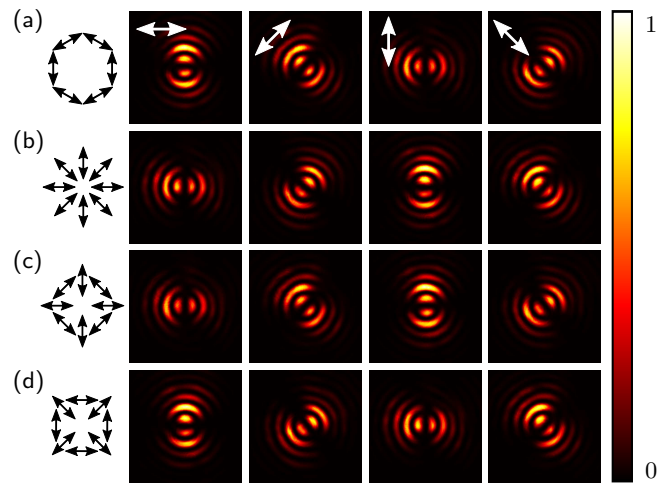  
\caption{Examples of non-separable vBG modes with respective polarisation analysis. The normalised intensity distribution in the $z_0$-plane for different orientations of a polariser are shown. The respective orientation is indicated by white arrows in (a). The according polarisation distribution is highlighted by black arrows. \label{fig:Polarization}} 
\end{figure}

\noindent In order to investigate the relation between the self-healing of propagation invariant beams and the non-separability of light modes, we apply vector Bessel beams, or, more precisely, vector Bessel-Gaussian (vBG) modes. These modes are classically entangled in their spatial and polarisation degrees of freedom (DoF) as explained in the next section. As illustrated in Fig. \ref{fig:Concept}(d1), these beams are generated by a suitable combination of an SLM (SLM$_1$), half wave plates ($\frac{\lambda}{2}$) and a $q$-plate (q), a device capable of correlating the polarisation and spatial DoFs\,\cite{Marrucci2006}. First, we create the fundamental scalar BG mode (linearly polarised in the horizontal direction) by encoding the binary Bessel hologram (Fig. \ref{fig:Concept}(b2)) on SLM$_1$. Within a $4f$-system we filter the first diffraction order with an aperture (A). If we now position, for example, a half wave plate, whose fast axis is oriented in a $45^{\circ}$ ($\pi/4$) angle with respect to the incoming horizontal polarisation, in combination with a $q$-plate ($q=1/2$) in the beam path, an azimuthally polarised vBG mode is created in the image plane of SLM$_1$ (see Fig. \ref{fig:Concept}(d1)). The desired mode is generated from the $q$-plate by coupling the polarisation DoF with the orbital angular momentum (OAM) via a geometric phase control, imprinting an OAM charge of $\pm 2q$ per circular polarisation basis to the passing beam\,\cite{Marrucci2006}. Note that wave and $q$-plate(s) do not need to be placed within the non-diffracting distance as we work in the paraxial regime ($\sin \alpha \approx \alpha$). The plates could even be located within the Fourier plane of SLM$_1$\,\cite{Milione2015d}. Moreover, we are of course not limited to azimuthally polarised vBG modes. Depending on the chosen number and orientation of wave plates, different polarisation structures are accessible\,\cite{Cardano2012} as depicted in Fig. \ref{fig:Polarization}. Here, we demonstrate the intensity distribution of different vBG beams in the $z_0$-plane analysed by a polariser (orientation indicated by white arrows).\\ 

In Fig. \ref{fig:Concept}(e1) we present the experimentally measured propagation invariant properties of these modes by the example of the azimuthal vBG beam (cf. Fig. \ref{fig:Polarization}(a)). The transverse intensity profile is shown for different positions $z\in [z_0,\,z_{\text{max}}]$. Consider that the outer rings of the vBG mode disappear with increasing $z$ due to the rhombus shape of the non-diffracting region.\\
As explained above, SLM$_1$ enables the inclusion of an obstruction within the holographically created scalar BG field. Following this, we are also able to apply the SLM for imparting an obstruction within the vBG mode: As SLM$_{1}$ is imaged by a $4f$-system to the $z_0$-plane of the vBG beam, an obstacle created by SLM$_1$ is also imaged to the $z_0$-plane of formed vector mode (cf. Fig.~\ref{fig:Concept}(d1)). As an example, we investigated the propagation properties of the azimuthal vBG mode if obstructed by an absorbing object with $R=200\,\mbox{\textmu m}$ created by SLM$_1$. Here, we included an additional horizontally oriented polariser to analyse the polarisation properties simultaneously. Results are shown in Fig. \ref{fig:Concept}(e2). For the programmed obstacle we calculate a self-healing distance of $z_{\text{min}} = 11.02\,\mbox{cm}$. The shown intensity distributions reveal a self-reconstruction of the beam including its polarisation properties after approximately $2z_{\text{min}}$, as expected.


\section{Non-separability of self-healing beams} 
\noindent
We define the non-separability of a self-healing vBG beam in the framework of quantum mechanics. This can be easily understood through the formal definition of the non-separability or entanglement of quantum systems; two systems A and B are separable if they can be written as a factorisable product of the two subsystems, i.e, $\ket{\Psi_{AB}}=\ket{A}\otimes\ket{B}$, conversely, the systems can be entangled if ($\ket{\Psi_{AB}} \neq \ket{A}\otimes\ket{B}$). Here, the two subsystems are analogously replaced with the internal DoF of the photons in the vBG beam. Importantly, a non-separable vector mode has maximally entangled polarisation and spatial components.\\
To quantify the non-separability between the polarisation and spatial components of the field, we employ simple measures borrowed from quantum mechanics  \cite{McLaren2015,Ndagano2016}. Consider an arbitrary (vBG) field with each photon described by the following state  
\begin{equation}
\ket{\Psi}_{k_r \ell} = \cos(\theta)\ket{u_{k_r,  \ell}} \ket{R} + \sin(\theta)\ket{u_{k_r,  -\ell}} \ket{L}, \label{eq: Field}
\end{equation}
\noindent where $\ket{R}$ and $\ket{L}$ are the canonical right and left circular polarisation states spanning the qubit Hilbert space $\mathcal{H}_2$. The infinite dimensional state vectors  $\ket{u_{k_r, \pm\ell}} \in \mathcal{H_\infty}$ represent the self-healing transverse eigenstates, namely scalar Bessel or BG modes of light (cf. Eq. (\ref{eq:BGmodes})), characterised by the continuous radial wave number $k_r$ and the topological charge $\ell$. The parameter $\theta$ determines whether $\ket{\Psi}_{k_r \ell}$  is purely vector (non-separable; $\theta=(2n+1)\pi/4,\, n\in \mathbf{Z})$, scalar (separable; $\theta=n\pi/2,\,\, n\in \mathbf{Z})$, or some intermediate state.

\begin{figure}[t]
	\centering
	\def\svgwidth{1.0\linewidth}\sffamily
	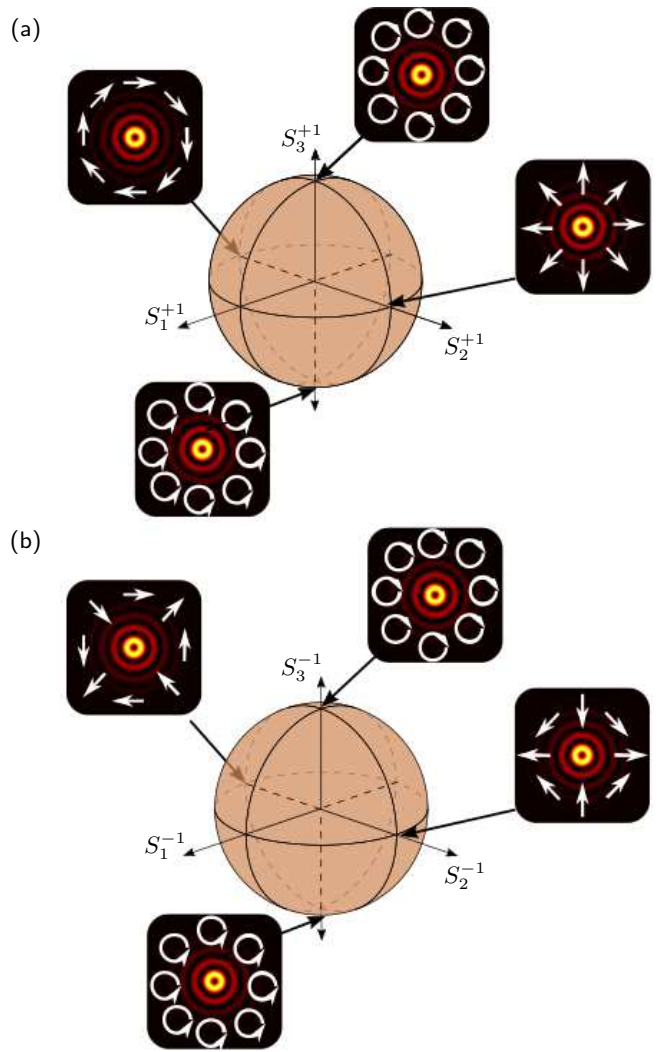
	\caption{Higher-order Poincar\'e spheres of self-healing BG beams spanned by the Stokes parameters (a) $S^{+1}_{1,2,3}$ for $\mathcal{H}_{2,\sigma,\ell=1}$ and (b) $S^{-1}_{1,2,3}$ for $\mathcal{H}_{2,\sigma,\ell=-1}$. The poles represent separable BG states with circular polarisation and $\ell=\pm1$, while non-separable vBG states are found on the equator (i.e. the plane intersecting the $S^{\pm1}_{1,2}$ plane).}	
	\label{fig:HOPS}
\end{figure}

The ``vectorness" of a given classical state can be determined via a vector quality analysis which is equivalent to measuring the concurrence of a quantum state $C$ \cite{Wootters2001}. Mathematically, the respective vector quality factor (VQF) has the form
\begin{equation}
\text{VQF}=\text{Re}(C)=\sqrt{1-s^2}=|\sin(2\theta)|. \label{eq:VQFpure}
\end{equation}
\noindent Here, $s$ is the Bloch vector defined as $s=\sum_{i}^{3}\langle \sigma_i\rangle$ with $\sigma_i$, $i=\lbrace 1,2,3\rbrace$, being the traceless Pauli operators spanning the so-called higher-order Poincar\'{e}  sphere (HOPS, see Fig. \ref{fig:HOPS})\,\cite{Millione2011}. The VQF takes values in the interval $[0,1]$, with 0 corresponding to a separable scalar mode and  1 representing a maximally non-separable vector mode. We employ the VQF measure as a figure of merit for determining the non-separability of partially obstructed and consequently self-healing vBG modes.

\subsection{Vector quality factor of obstructed beams}

\noindent
From wave optics, it is well-known that an obstructed beam can be modified in both phase and amplitude due to the interaction between the beam and the outer edges of an obstruction, resulting in diffraction. At the single photon level, this can be understood as modal coupling\,\cite{Sorelli2018}. That is, if some field with a transverse profile given by $\ket{ u_{k_{r_1}, \ell_1} }$ interacts with an obstruction, the state evolves following the mapping 
\begin{equation}
\ket{ u_{k_{r_1},\ell_1 }} \rightarrow \int\sum_{\ell} \alpha_{\ell}(k_r) \ket{ u_{{k_r},\ell}} \text{d}k_r, \label{eq. Diff}
\end{equation}
where the input mode spreads over all eigenmodes $\ket{u_{{k_r},\ell}}$. The coefficients $|\alpha_{\ell}(k_r)|^2$ represent the probability of the state $\ket{ u_{k_{r_1  },\ell_1}}$ scattering into the eigenstates $\ket{u_{{k_r},\ell}}$ with the property that $\int\sum_{\ell}|\alpha_{\ell}(k_r)|^2\text{d}k_r=1$. Applying the mapping of Eq. (\ref{eq. Diff}) to the scattering of the self-healing vBG mode presented in Eq. (\ref{eq: Field}) and post selecting a particular $k_r$ and $\ell$ values yields the state 
\begin{eqnarray}
\ket{\Phi_{k_r, \ell}}=& a \ket{ u_{{k_r},\ell}}\ket{R} + b \ket{ u_{{k_r},-\ell}}\ket{R}\nonumber\\ 
&+ c \ket{ u_{{k_r},\ell}}\ket{L} + d \ket{u_{k_r,\ell}}\ket{L}, \label{post}
\end{eqnarray}
where $|a|^2 +|b|^2+|c|^2+|d|^2=1$. Consequently, $\ket{\Phi_{k_r, \ell}}$ restricts the measurement of the photons to the four-dimensional Hilbert space  $\mathcal{H}_4=\text{span}\big(\{\ket{R}, \ket{L}\}\otimes\{\ket{ u_{{k_r},\ell}}, \ket{ u_{{k_r},-\ell}}\}\big)$ which can be written as the direct sum
\begin{equation}
	\mathcal{H}_4= \mathcal{H}_{2,\sigma,\ell} \oplus \mathcal{H}_{2,\sigma,-\ell}.\label{eq:Hilbert}
\end{equation}
\noindent The subspaces $\mathcal{H}_{2,\sigma,\pm\ell}=\text{span}(\ket{u_{k_r, \pm \ell}}\ket{R}, \ket{u_{k_r,\mp\ell}}\ket{L})$ are topological unit spheres for spin-orbit coupled beams belonging to the family of HOPS\,\cite{Millione2011}. \\
\noindent Invoking the equivalence between VQF and concurrence enables us to exploit the definition of concurrence with $C=\sqrt{1-\text{Tr}(\rho_{\psi}^2)}$ for a given density matrix $\rho_\psi=\ket{\psi}\bra{\psi}$. In quantum mechanics, if the state of a two qubit system written in the logical computation basis is given by the pure state
\begin{equation}
 \ket{\Psi} = c_1\ket{0}\ket{0}+c_2\ket{0}\ket{1}+c_3\ket{1}\ket{0}+c_4\ket{1}\ket{1},
\end{equation}
\noindent satisfying the normalization condition $\sum_{j=1}^4 |c_{j}|^2=1$, the concurrence is $C=2|c_1c_4-c_3c_2|$. By replacing the two qubit logical basis with the $\mathcal{H}_4$ basis from the two HOPSs, we can equivalently write the VQF as 
\begin{equation}
\text{VQF}=2|ad-cb|, \label{VQF_VM}
\end{equation}
\noindent following Eq. (\ref{post}).\\
As an illustrative example, consider the azimuthally polarised vBG mode given by 
\begin{equation}
\ket{\psi}_{k_r,1} =\frac{1}{\sqrt{2}}\big( 
 \ket{u_{k_r, 1}} \ket{R} - \ket{u_{k_r,  -1}} \ket{L}\big). \label{eq:Fieldradial}
\end{equation}

\noindent Upon diffracting off the edges of an obstruction, one expects the mode coupling profiled in Eq. (\ref{post}) to occur (by restricting $|\ell|=1$). However, since the radial profile of BG modes enables the transverse structure to self-heal, $cb\rightarrow0$ and $ad\rightarrow-\frac{1}{2}$, and therefore $\text{VQF}\rightarrow 1$ giving rise to the self-healing of the non-separability. This behavior is predicated to occur best after twice of the minimum self-healing distance $z_{\text{min}}$.\\

\noindent To experimentally quantify the characteristics of vectorness, i.e. classical entanglement, in relation to the self-healing properties of vGB modes, we apply a configuration as indicated in Fig. \ref{fig:Concept}(d2). By this configuration, consisting of a quarter wave plate ($\frac{\lambda}{n}$, $n=4$), a polarisation sensitive SLM (SLM$_2$), a Fourier lens (L$_{\mathcal{F}}$) and a CCD camera, we determine the expectation values of the Pauli operators $\langle\sigma_{i}\rangle$, $i = \lbrace 1,\,2,\,3\rbrace$ by 12 on-axis intensity measurements or six identical measurements for two different basis states\,\cite{McLaren2015,Ndagano2016}. These values are used to determine the VQF according to Eq. (\ref{eq:VQFpure}). \\
If circular polarisation is chosen as basis ($\ket{R},\,\ket{L}$), the projection measurements represent two OAM modes of topological charge $\ell$ and $-\ell$, namely $\ket{u_{k_r,\pm\ell}} = E_{\pm\ell}^{\text{BG}}$, as well as four superposition states $\ket{u_{k_r,\ell}} + \exp(\text{i}\gamma) \ket{u_{k_r,-\ell}}$ with $\gamma = \lbrace 0, \pi \ell /2, \pi \ell, 3\pi\ell/2 \rbrace$. As we use a $q$-plate with $q=1/2$, our measurements are performed for $\ell = 1$. Following Tab. \ref{tab:Tomography}, we calculate expectation values $\langle\sigma_i \rangle$ from
\begin{align*}
&\langle\sigma_1\rangle = I_{13}+I_{23}-(I_{15}+I_{25}),\\
&\langle\sigma_2\rangle = I_{14}+I_{24}-(I_{16}+I_{26}),\\
&\langle\sigma_3\rangle = I_{11}+I_{21}-(I_{12}+I_{22}).
\end{align*}
On-axis intensity values $I_{uv}$, $u = \lbrace1,2\rbrace$ , $v = \lbrace1,2,...,6\rbrace$, are normalised by $I_{11}+I_{12}+I_{21}+I_{22}$. The respective polarisation projections are performed by inserting a quarter wave plate, set to $\pm 45^{\circ}$, in combination with the polarisation selective SLM$_2$. Further, SLM$_2$ is responsible for the OAM projections. For this purpose, we encode the OAM as well as superposition states as phase-only holograms according to the binary Bessel function in Eq.~(\ref{eq:BinaryBessel}). Finally, the on-axis intensity is measured in the focal plane of a Fourier lens by means of a CCD camera.\\
Crucially, the decoding SLM$_2$ is placed at a adequately chosen distance $\Delta z = 23\,\mbox{cm}$ from the $z_0$-plane so that we are able to access different levels of self-healing without the need to move the detection system. That is, by changing the radius $R$ of the digitally created obstruction, the vBG mode can fully self-heal in front of ($2z_{\text{min}}< \Delta z$) or behind ($2z_{\text{min}}> \Delta z$) this SLM$_2$, as indicated in Fig.~\ref{fig:Concept}(d2) by the black shadow region or the green and yellow dashed lines, respectively. 
  \begin{table}[h]
 \caption{Normalised intensity measurements $I_{uv}$ for the determination of expectation values $\langle \sigma_i \rangle$. \label{tab:Tomography}}
 \begin{ruledtabular}
 \begin{tabular}{c|c c c c c c c}
Basis states & $l=1$ & $-1$ & $\gamma =0$ & $\pi/2$ & $\pi$ & $3\pi/2$& \\ \hline \hline
Left circular $\vert L\rangle$& $I_{11}$ & $I_{12}$ & $I_{13}$ & $I_{14}$ & $I_{15}$ & $I_{16}$ &\\ 
Right circular $\vert R\rangle$& $I_{21}$ & $I_{22}$ & $I_{23}$ & $I_{24}$ & $I_{25}$ & $I_{26}$& \\ 
 \end{tabular}
 \end{ruledtabular}
 \end{table}

\begin{figure*}[tb]
\centering
\def\svgwidth{1.0\linewidth}\sffamily
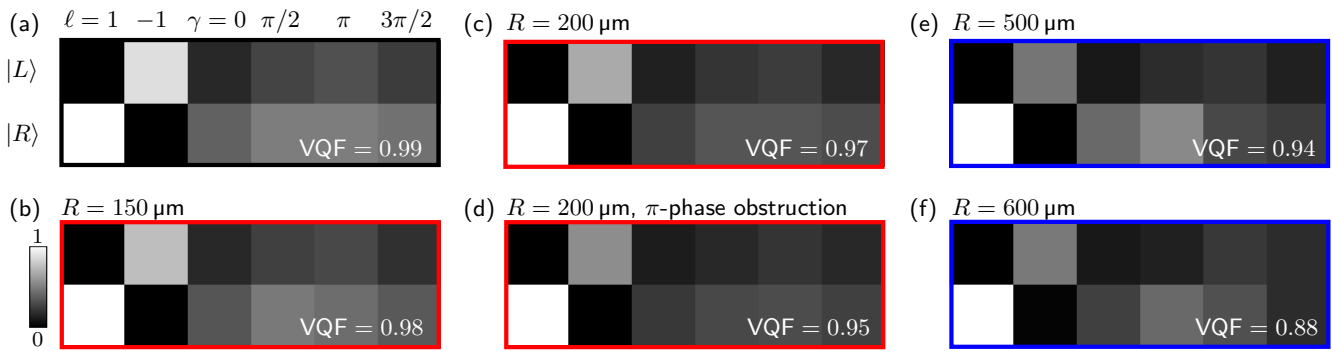  
\caption{Vector quality analysis of self-healing vBG mode (azimuthally polarised) without obstacle (a), with absorbing obstacles of radius (b) $R = 150\,\mbox{\textmu m}$, (c) $R=200\,\mbox{\textmu m}$, (e) $R = 500\,\mbox{\textmu m}$, and (f) $R=600\,\mbox{\textmu m}$, as well as (d) phase obstacle (homogeneous phase shift of $\pi$) of radius $R=200\,\mbox{\textmu m}$. Results for beams which (do not) fully self-heal before the analysis are marked (blue) red with respective VQF within the measured normalized matrices of on-axis intensity values arranged according to Table.~\ref{tab:Tomography}.  \label{fig:StateTomography}} 
\end{figure*}

\subsection{Experimental quantification of vectorness}
\noindent First, we prove that pure undisturbed vBG modes show the maximum degree of non-separability. This is exemplified by the analysis of an azimuthally polarised vBG beam. The experimentally measured on-axis intensity values are visualised in Fig. \ref{fig:StateTomography}(a), arranged according to Tab. \ref{tab:Tomography}. The measurements result in a VQF of $0.99$, verifying our expectation. In a next step, we digitally impart different obstacles and determine the respective VQF. On the one hand, we chose obstacles allowing the beam to self-reconstruct within $\Delta z =23\,\mbox{cm}$, namely absorbing obstructions with $R=150\,\mbox{\textmu m}$ ($2z_{\text{min}} = 2\cdot 8.27\,\mbox{cm} = 16.54\,\mbox{cm}$) or $R=200\,\mbox{\textmu m}$ ($2z_{\text{min}} = 2\cdot 11.03\,\mbox{cm} = 22.06\,\mbox{cm}$) and a phase obstacle with $R=200\,\mbox{\textmu m}$ creating a homogeneous phase shift of $\pi$ (Fig. \ref{fig:StateTomography}(b)-(d)). On the other hand, we program absorbing obstructions with $R = 500\,\mbox{\textmu m}$ ($2z_{\text{min}} = 2\cdot 27.57\,\mbox{cm} = 55.14\,\mbox{cm}$) and $R = 600\,\mbox{\textmu m}$ ($2z_{\text{min}} = 2\cdot 33.03\,\mbox{cm} = 66.06\,\mbox{cm} $), for which $2z_{\text{min}}>\Delta z$ and even $z_{\text{min}}>\Delta z$ (Fig. \ref{fig:StateTomography}(e), (f)). The respective measured VQFs are shown within each subfigure. \\
Obviously, the degree of non-separability, i.e. the VQF, decreases with increasing absorbing obstacle size $R$ (Fig. \ref{fig:StateTomography}). For self-healed beams with absorbing obstacles (b)-(c), the VQF differs only minimally from the non-obstructed case (a). Note that a phase obstruction of the same radius (d) results in a slightly larger deviation. In contrast to absorbing obstacles, causing a loss of information, phase obstructions do not cut but vary information. Since in the case of absorbing obstacles the cut information is also included within the passing plane waves, the loss can be compensated within $2z_{\text{min}}$. In the phase obstruction case, the varied information represent additional information, i.e. noise, within the non-diffracting beam, which is not eliminated when being decoded by the SLM. Hence, the beam stays disturbed and, as a consequence, the VQF decreases.\\
Further, if the beam cannot fully reconstruct before being decoded by SLM$_2$, thus, if we analyse the degree of entanglement within the self-healing distance (Fig. \ref{fig:StateTomography}(e), (f)), the decrease in VQF is relatively large in comparison to the self-healed versions in Fig. \ref{fig:StateTomography}(b), (c). However, note that in both cases, the self-reconstructed and non-reconstructed beams, the VQF is $\geq 0.88$. Consequently, the beams are closer to being vector or non-separable (VQF $=1$) than scalar or separable (VQF $=0$). This is due to the fact that there is always undisturbed information reaching the SLM$_2$, and only little scattering into other modes, i.e., little modal coupling. We thus may conclude that the beam is always non-separable, but due to noise caused by the obstacle the measurement shows deviations from pure non-separability. As the noise is annihilated with propagation distance (for absorbing obstacles), the quantitative value for non-separability, namely VQF, recovers. This effect is similar to what we call ``self-healing" in the case of amplitude, phase and polarisation of vBG modes: Obstacles add noise to the information on these degrees of freedom. Upon propagation, these perturbations vanish and the pure vBG mode information is left so that the beam and its properties seem to self-reconstruct.\\
Consequently, we conclude that the lower the level of self-healing, i.e. the larger the obstacle, the smaller the VQF. This means, not only amplitude, phase and polarisation properties of the vBG mode reconstruct with distance behind an obstruction, but also the degree of non-separability. \\

\begin{figure*}[tb]
\centering
\def\svgwidth{1.0\linewidth}\sffamily
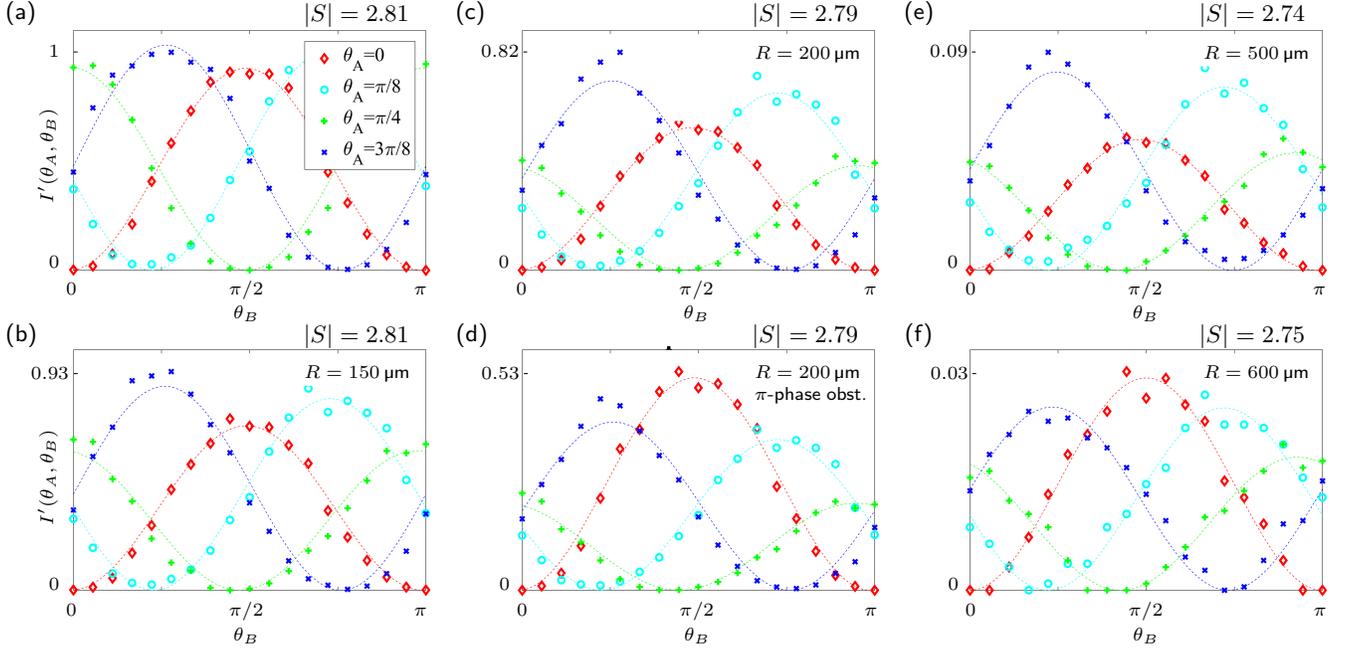  
\caption{Bell-type curves for azimuthal vBG mode with four different orientations $2\theta_A$  of the half wave plate to determine the Bell parameter $|S|$. The investigations were performed undisturbed (a) as well as for differently sized obstacles (radii $R=\lbrace 150,\,200,\,500,\,600\rbrace\,\mbox{\textmu m}$ in (b)-(f)) and a $\pi$-phase obstacle ($R=200\,\mbox{\textmu m}$) (d). Shown on-axis intensity measurements $I'(\theta_A, \theta_B)$ for (a)-(f) are normalised according to the maximum intensity measured without obstacle (a). Dashed curves represent $\cos^2$-fits used to determine $S$. \label{fig:BellAnalysis}} 
\end{figure*}

\subsection{CHSH Bell-like inequality violation}
\noindent To confirm our results with respect to the VQF, we performed an additional investigation of the degree of entanglement or non-separability using the Bell parameter\,\cite{McLaren2015}. More specifically, we perform a Clauser-Horne-Shimony-Holt (CHSH) inequality measurement\,\cite{clauser1969}, the most commonly used Bell-like inequality for optical systems, to demonstrate the degree of entanglement between polarisation and spatial DoFs. Instead of measuring a single DoF, e.g. polarisation or OAM, non-locally, we analyse two DoF locally on the same classical light field\,\cite{McLaren2015}. For this purpose, we placed a half wave plate ($\frac{\lambda}{n}$, $n=2$) in front of SLM$_2$ (see Fig. \ref{fig:Concept}(d2)) and measured the on-axis intensity $I(\theta_A,\theta_B)$ for different angles $2\theta_A = \lbrace 0,\, \pi/8,\,\pi/4,\,3\pi/4\rbrace$ of the half wave plate. Here, $\theta_B \in [0,\,\pi]$ represents the rotation angle of the hologram encoded on SLM$_2$ by $\ket{u_{k_r,\ell}} + \exp(\text{i}2\theta_B) \ket{u_{k_r,-\ell}}$ ($\ell = 1$, $k_r = 18\,\mbox{rad\,mm}^{-1}$). \\
We define the CHSH-Bell parameter $S$ as
\begin{equation}
S = E(\theta_A,\theta_B)-E(\theta_A,\theta_B')+E(\theta_A',\theta_B)+E(\theta_A',\theta_B'), \label{eq:BellParameter}
\end{equation}
with $E(\theta_A,\theta_B)$ being calculated from measured on-axis intensity according to
\begin{align}
& E(\theta_A,\theta_B) = \frac{A(\theta_A,\theta_B)-B(\theta_A,\theta_B)}{A(\theta_A,\theta_B)+B(\theta_A,\theta_B)}.\\
& A(\theta_A, \theta_B) = I(\theta_A,\theta_B)+I\left(\theta_A+\frac{\pi}{2},\theta_B+\frac{\pi}{2}\right),\nonumber\\
& B(\theta_A, \theta_B) = I\left(\theta_A+\frac{\pi}{2},\theta_B\right)+I\left(\theta_A,\theta_B+\frac{\pi}{2}\right).\nonumber
\label{eq:BellParameterE}
\end{align}
The values of $S$ ranges from $|S|\leq 2$ for separable states, up to $|S|= 2\sqrt{2}$ for entangled or non-separable states. Our experimental results with according Bell parameters $|S|$ are presented in Fig. \ref{fig:BellAnalysis}. As before, we performed our investigation without obstacle (a) as well as with different absorbing obstacles of radii $R = \lbrace 150,\, 200,\, 500,\, 600\rbrace\,\mbox{\textmu m}$ (b)-(f), or (d) phase-obstructing. Note that the intensity values are normalised with respect to the maximum value measured in the unobstructed case ($I'(\theta_A, \theta_B)$). Obviously, the ratio of measurable maximum intensity ($I'_{\text{max}}(R)$) decreases dramatically depending on the size of the obstruction, as it can be seen on the $I'(\theta_A, \theta_B)$-axis (vertical) of Fig. \ref{fig:BellAnalysis}. However, the Bell parameter does not change significantly if we measure in the fully reconstructed regime ($2z_{\text{min}}<\Delta z$ for (b)-(d)) compared to the undisturbed vBG mode (a). In accordance with our VQF analysis, $|S|$ reveals bigger changes if the beam is not fully self-healed when it is analysed (e), (f). In total, even if the intensity lowers to some percentage (f) of the original maximum value (a), all measurements validate a violation of the Bell inequality, matching our non-separability analysis results based on vectorness. 

\subsection{The relation of self-healing and non-separability}
 \begin{table}[h]
 \caption{Quantification of non-separability properties of self-reconstructing vBG modes as function of the self-healing level, given by the obstacle radius $R$.  \label{tab:Summarize}}
 \begin{ruledtabular}
 \begin{tabular}{c|c c c c c c c}
$R$ in \textmu m & $0$ & $150$ & $200$ & $200$, $\pi$-obst. & $500$ & $600$& \\ \hline \hline
VQF & $0.99$ & $0.98$ & $0.97$ & \textcolor{gray}{$0.95$} & $0.94$ & $0.88$ &\\ 
$|S|$ & $2.81$ & $2.81$ & $2.79$ & \textcolor{gray}{$2.79$} & $2.74$ & $2.75$& \\ 
$I'_{\text{max}}$ & $1$ & $0.93$ & $0.82$ & \textcolor{gray}{$0.53$} & $0.09$ & $0.03$ & \\
 \end{tabular}
 \end{ruledtabular}
 \end{table}
 
\noindent In Table \ref{tab:Summarize} we summarize our results emphasizing the dependence of the VQF, the maximum intensity $I'_{\text{max}}$ as well as Bell parameter $|S|$ on the size $R$ of included obstructions, i.e. on the self-healing level. The maximum intensity $I'_{\text{max}}(R)$ reveals an approximately Gaussian decrease with increasing obstacle size which reflects the Gaussian envelope of investigated vBG mode. Simultaneously to the intensity, the VQF as well as $|S|$ decrease as demonstrated in previous sections. However, only small changes are observed as only minor noise is disturbing vBG modes if obstructions are included. In short, both, vector quality as well as Bell analysis reveal a similar behavior of non-separability with respect to changes in the obstruction size, demonstrating the self-healing of the degree of entanglement within obstructed vBG beams.

\section{Conclusion and Discussion}

\noindent It is well-known that the phase and amplitude of vector Bessel beams self-heal in the presence of an obstacle that partially blocks its path. However, there are no reports about the effect of this on the coupling between phase and polarisation, known as classical entanglement. In this work, we presented for the first time to our knowledge experimental evidence that even though the coupling between these two degrees of freedom decreases after passing an obstruction, it eventually restores itself to its maximum value. For this purpose, we dynamically realized vector Bessel Gaussian (vBG) modes with digital obstructions by combining holography-based generation of structured light with a q-plate. We quantified the degree of non-separability between the spatial shape and polarisation using two different means: the vector quality factor (VQF) and a classical version of the CHSH Bell-like inequality. By a specific design of our detection system combined with digital variation of the obstruction size, different levels of self-healing were accessible, which enabled the relation between degree of non-separability and self-healing level to be analysed.
The measured VQF values showed that the degree of classical entanglement increases as function of decreasing object size. Analogously, the measured CHSH $S$ parameter values show a similar dependence on the objects size or self-healing level of the vBG beam, showing in all cases a clear violation of the CHSH inequality. This behavior can be interpreted as a self-healing in the degree of non-separability as function of the distance from an obstacle since a decrease in obstacle size is comparable to placing the detector further from the object.\\
The complex fields used in this study were separable in radial profile, defined by the $k_r$ vector of the conical waves, but non-separable in the azimuthal profile and polarisation, the latter being defined by the OAM of each polarisation component. Intriguingly, it is the radial profile that leads to the self-healing of the non-separability, despite itself being separable. Thus the local entanglement in angular momentum (spin and OAM) can be made more resilient to decay from obstructions by engineering the unused degree of freedom in a judicious manner.
  
\newpage
%

\end{document}